\documentclass[11pt,letterpaper,english]{revtex4}
\usepackage{times}
\usepackage[T1]{fontenc}
\usepackage[latin1]{inputenc}
\usepackage{subfigure}
\usepackage{amsmath}
\usepackage{graphicx}

\makeatletter

\usepackage{babel}
\makeatother
\begin{document}

\title{Self consistent determination of plasmonic resonances in ternary nanocomposites}

\author{Hernando Garcia}

\email{hgarcia@siue.edu}

\affiliation{Department of Physics, Southern Illinois University, Edwardsville, IL 62026}

\author{Justin Trice}

\affiliation{Department of Physics, Washington University in St. Louis, MO 63130}

\affiliation{Center for Materials Innovation, Washington University in St. Louis, MO 63130}

\author{Ramki Kalyanaraman}

\email{ramkik@wuphys.wustl.edu}

\affiliation{Department of Physics, Washington University in St. Louis, MO 63130}

\affiliation{Center for Materials Innovation, Washington University in St. Louis, MO 63130}

\author{Radhakrishna Sureshkumar}

\email{suresh@che.wustl.edu}

\affiliation{Department of Energy, Environmental and Chemical Engineering, Washington University in St. Louis,
MO, 63130}

\affiliation{Center for Materials Innovation, Washington University in St. Louis, MO 63130}

\begin{abstract}
We have developed a self consistent technique to predict the behavior of plasmon resonances in
multi-component systems as a function of wavelength. This approach, based on the tight lower
bounds of the Bergman-Milton formulation, is able to predict experimental optical data, including
the positions, shifts and shapes of plasmonic peaks in ternary nanocomposites without using any
fitting parameters. Our approach is based on viewing the mixing of 3 components as the mixing
of 2 binary mixtures, each in the same host. We obtained excellent predictions of the experimental
optical behavior for mixtures of Ag:Cu:SiO$_{\text{2}}$ and alloys of Au-Cu:SiO$_{\text{2}}$
and Ag-Au:H$_{\text{2}}$O, suggesting that the essential physics of plasmonic behavior is captured
by this approach.
\end{abstract}
\maketitle
Accurate control and prediction of plasmonic behavior in metal nanoparticles in various configurations
are expected to realize future ultra-high density integrated photonic devices \citep{Koehl05}.
For example, plasmonic effects in metal nanoparticle waveguide arrays and metal-dielectric interfaces
have been used to transport electromagnetic energy below diffraction limited dimensions \citep{karalis05,maier03b,maier03,krenn99,quinten98}.
Glasses containing metal nanoparticles exhibit intensity dependent refractive index values several
orders of magnitude greater than that of silica glass due to dielectric and quantum confinement
effects \citep{SchmittRinkPhysRevB87}. These enhanced effects, occurring due to surface plasmon
effects at the interface of the metal and the dielectric, could allow various use of linear and
nonlinear optical responses of mixtures \citep{lopez04,YangApplPhysA96,WangAppPhyLett83}, provided
appropriate materials are selected \citep{garcia06b}. Experimental observations have also demonstrated
that composites containing multiple metals show multiple plasmonic peaks as well as dependence
of such peaks on metal fraction \citep{MagruderNonCrystSold94,BattaglinNucInsMethPhysB00}. Therefore,
the large optical nonlinearities, the waveguiding capability and multiple-wavelength sensitivity
offered by plasmon behavior makes it an important area of study with potential applications to
optoelectronics, sensing, etc. \emph{In this Letter, we develop and demonstrate a self consistent
technique to accurately predict the effective dielectric constant of multi-component systems
and apply it to ternary nanocomposites for which reliable experimental data are available} \emph{\citep{MagruderNonCrystSold94,BattaglinNucInsMethPhysB00,MoskovitsJChemPhys02}.}

The accurate prediction of the optical properties of nanocomposites made from multiple optical
materials is an outstanding problem in current research. This is primarily because the effective
permittivity is not uniquely determined by the optical properties of the individual components
but also requires knowledge of the composite microstructure \citep{Bergman80,MiltonJAppPhys1981}.
Previous theoretical work in this area has focused primarily on estimating the effective permittivity
of two-component mixtures where one of the components is a dielectric and the other is a metal
\citep{SihvolaIEEE99}. There are numerous models that predict the effective permittivity of
\emph{two} component composites \citep{GarnettTranRoySoc1904,BruggemanAnnPhys1935,Kohler81}.
Moreover, there are already established rigorous bound for the values of the permittivity in
such composites \citep{HashinJApPhy62,BergmanPhysRep1981,MiltonJAppPhys1981}. However, these
models have not been extended to or applied towards interpreting the behavior of multi-metal
composites and typically multi-parameter fitting routines have been used to reproduce experimental
data \citep{anderson98,MoskovitsJChemPhys02}. But such fitting approaches are of limited use
in predicting the fundamental behavior of unexplored nanocomposite systems and hence, are not
generally applicable tools for new product design. Recently it was shown that the problem of
calculating the bounds for a multicomponent system can be reduced to the determination of the
convex-hull generated in the complex plane when the volume fraction of each component act as
a barycentric coordinate for the hull \citep{PeiponenOptLett06}. One of the limitations of this
approach is that it can be used only for obtaining the loosest bounds which are the Wiener bounds
\citep{Wiener12}. These loose bounds cannot predict the location of the plasmon resonances in
frequency space for a simple two component system. This is because the resonance enhancement
of the absorption takes place at $f_{m}(\epsilon_{h}-\epsilon_{1})+\epsilon_{1}=0$ and is in
contradiction to the well known condition for the resonance at $2\epsilon_{h}+\epsilon_{1}=0$,
where $f_{m}$ is the volume fraction of the metal, $\epsilon_{h}$ is the dielectric permittivity
of the host and $\epsilon_{1}$ is the real part of the dielectric permittivity of the metal
or inclusion. 

Here we report a self consistent technique that accurately predicts the plasmonic behavior, including
peak positions, shifts and shapes, in multi-metal mixtures and/or alloy nanocomposites without
the need for any fitting parameters. Our approach begins with the fact that for a two component
nanocomposite the tightest lower bound on the dielectric permittivity can be expressed through
the Bergman-Milton formula \citep{Bergman80,MiltonJAppPhys1981}, (which is based on restricting
the values of the effective dielectric permittivity in the complex plane using suitably defined
conformal mappings) as: \begin{equation}
\epsilon_{eff}(\gamma)=\left[\frac{f_{a}}{\epsilon_{a}}+\frac{f_{h}}{\epsilon_{h}}-\frac{2f_{a}f_{h}\left(\epsilon_{a}-\epsilon_{h}\right)^{2}}{3\epsilon_{a}\epsilon_{h}\left[\epsilon_{h}\gamma+\epsilon_{a}\left(1-\gamma\right)\right]}\right]^{-1}\label{eq:BergMilton}\end{equation}
where $f_{a}$ and $f_{h}$ denote the volume fractions of the constituent materials with relative
permittivity $\epsilon_{a}$ and $\epsilon_{h}$ respectively, and $f_{a}+f_{h}=1$. The parameter
$\gamma$ takes the values $2(1-f_{a})/3\leq\gamma\leq1-2f_{a}/3$. In this work we use $\gamma=1-2f_{a}/3$
because it corresponds to the exact Maxwell Garnett mixing rule \citep{garnett1904}. The upper
bound corresponds to the complementary mixture and is not of interest to us. 

The central idea of this paper is to use Eq. \ref{eq:BergMilton} in a self consistent fashion
within the effective medium approximation to calculate the effective permittivity of a three
component or ternary nanocomposite. We begin with the hypothesis that a three component system
formed by $\epsilon_{a}$, $\epsilon_{b}$ and the host $\epsilon_{h}$ can be viewed as a mixture
of two components, each having an effective permittivity $\epsilon_{eff}^{a,h}$ and $\epsilon_{eff}^{b,h}$
calculated using Eq. \ref{eq:BergMilton}. In this mixing process the average electric field
within the composite is held fixed at every stage of mixing. The constraint in the volume fraction
is introduced while calculating the individual effective permittivities while the final effective
permittivity is calculated using \emph{equal} volumes of each mixture. Moreover, notice that
$\epsilon_{h}$, which represents the dielectric permittivity of the host matrix, is common to
both mixtures. This simple binary mixing rule, depicted graphically in Fig. \ref{fig:Schematic-of-mixing},
can be stated as follows: \emph{the effective permittivity of an N-component mixture can be determined
by mixing N-1 binary mixtures, each comprising of a host and a distinct metal, with the host
being common to the N-1 pairs}. In this Letter we focus on applying this mixing rule to ternary
systems for which experimental data exists.

In order to test the validity of our hypothesis we have calculated the optical absorption coefficient
of ternary composites containing multi-metal and/or alloy nanocomposites using experimental data
for the individual permittivities of the metals obtained from ref. \citealp{SopraDataBase}.
The binary mixing rule was applied to the formation of nanocomposites made from mixtures of two
metals, as shown in Fig. \ref{fig:Schematic-of-mixing}(a), and for alloys, as shown in Fig.
\ref{fig:Schematic-of-mixing}(b). We also introduced a modification to the imaginary component
of the permittivity of each metal in order to account for the enhanced rate of electron scattering
due to particle size-dependent effects by modifying the Drude model \citep{mermin76}. This was
accomplished by expressing the imaginary component in the high frequency limit as follows: \begin{equation}
\epsilon_{2}=\frac{\omega_{p}^{3}}{\omega^{2}\tau_{eff}}=\epsilon_{2}^{bulk}\left(\frac{2d+\nu_{F}\tau_{bulk}}{2d}\right)\label{eq:epscorr}\end{equation}
where $\omega_{p}$ is the bulk plasmon resonance frequency of the metal, and $\tau_{eff}$ is
an effective relaxation time given by: \begin{equation}
\frac{1}{\tau_{eff}}=\frac{1}{\tau_{bulk}}+\frac{\nu_{F}}{2d}\label{eq:taueff}\end{equation}
 where $\tau_{bulk}$ is the bulk relaxation time of the electron, $\nu_{F}$ is the speed of
the electrons close to the Fermi surface, and $d$ is the nanoparticle diameter. The second term
in Eq. \ref{eq:taueff} takes into account the collision rate of the electron with the metal
walls for spherically shaped particles. 

In order to compare our predictions to previous experimental measurements, we identified multi-metal
composites for which experimental observations of the plasmonic behavior are available including
the three parameters required for the calculations, namely: (i) the average metal particle size;
(ii) the length of the sample which determines the absorption, and (iii) the composition of the
metal particles. Based on this, we compared theoretical predictions with the experiments of Magruder
and co-workers on \emph{Ag and Cu} \emph{mixtures} in SiO$_{\text{2}}$ \citep{MagruderNonCrystSold94},
by Battaglin and co-workers on \emph{Au-Cu alloys} in SiO$_{\text{2}}$ \citep{BattaglinNucInsMethPhysB00}
and by Moskovits and co-workers on \emph{colloidal solutions of Au-Ag alloy nanoparticles} in
H$_{\text{2}}$O \citep{MoskovitsJChemPhys02}. In our calculations, the optical absorption in
a silica glass or water matrix was estimated in terms of the optical density $(OD)$ given by
$OD=\alpha l/2.3$, where $\alpha$ is the absorption coefficient of the final mixture given
by $\alpha=\frac{2\pi}{n_{o}\lambda}Im(\epsilon_{eff}(\gamma))$, where $n_{o}$ is the dielectric
constant of the medium (corrected for addition of the nanoparticles), $\lambda$ is the incident
wavelength and and \emph{l} is the length of the sample. 

Magruder et al. \citep{MagruderNonCrystSold94} measured the behavior of the plasmonic peaks
as a function of the volume fractions of the individual metals in a composite of Ag and Cu in
SiO$_{\text{2}}$. Here, the various ratios of the metal ions were achieved by sequential ion
implantation. The resulting composites had particles with a mean size of \emph{d = 30} nm for
Ag and Cu in ratios of 9:3, 6:6, and 3:9 respectively. The central observations were that: (i)
a dominant peak, attributed to Ag, shifts from \textasciitilde{}410 nm to 440 nm as the Ag concentration
was increased; and (ii) a second weak peak near \textasciitilde{}575 nm appeared as the Cu concentration
increased. Using the above values for the particle diameter and individual metal concentrations
along with the experimental values for $\nu_{F}$ and $\tau_{bulk}$ for Cu and Ag we calculated
the effective permittivity using the mixing rule developed above and estimated the plasmonic
behavior of this system. Fig. \ref{fig:Magruder} compares the experimental data extracted from
ref. \citep{MagruderNonCrystSold94} (symbols) with the results of our predictions (lines). The
theory, without the use of any fitting parameters, predicts the position of the plasmon peaks
as well as the shift occurring due to change in Ag concentration. Our calculations predicted
a shift from 417 nm to 452 nm with increasing Ag concentration. The calculation also accurately
captured the appearance of the second peak at 572 nm with increasing Cu concentration. Moreover,
the widths of the individual peaks, which is extremely sensitive to particle size, were fairly
well represented by our model. The peak widths were a consequence of the correction introduced
to the electron scattering time based on the particle size. 

In Fig. \ref{fig:Battaglin} we compare model predictions with the data of Battaglin et al. \citep{BattaglinNucInsMethPhysB00}
in which the plasmonic behavior was studied as a function of alloy concentration and particle
size for Au-Cu alloys in SiO$_{\text{2}}$. Equal ratios of Au and Cu were implanted and then
subjected to different annealing conditions. The as-deposited sample contained an alloy with
Au$_{\text{2}}$Cu$_{\text{1}}$ composition (with the remaining Cu in atomic state) and average
particle size of $3.8\pm1.2\, nm$. The sample annealed in H$_{\text{2}}$ had a Au$_{\text{1}}$Cu$_{1}$
alloy with an average particle size of $8.7\pm2.5\, nm$ while the sample annealed in air contained
primarily Au nanoparticles of $33\pm15\, nm$ diameter, with the Cu preferentially found as an
oxide in the near surface of the samples. For the theoretical calculations we first created the
appropriate alloys using the mixing rule (depicted in Fig. \ref{fig:Schematic-of-mixing}(b))
and then the final composite based on the volume fraction of the alloy and the experimentally
assigned particle diameters. The results are shown in Fig. \ref{fig:Battaglin}, with experimental
data as symbols and theory as lines. Once again the theory accurately predicts the peak positions
and shift for this alloy system without the need for any fitting parameters. 

In Fig. \ref{fig:Moskovits} we compare the theoretically predicted and experimentally observed
behavior of colloidal solutions containing Au-Ag core-shell nanoparticle alloys of various compositions
in a water matrix, as measured by Moskovits and co-workers \citep{MoskovitsJChemPhys02}. Their
primary conclusions were: (i) a single plasmon peak appeared for the core-shell structures and
this peak shifted with changing Au-Ag concentration. This behavior was attributed to the formation
of an alloy in the shell whose composition changed with varying Au-Ag fraction; and (ii) the
trend could be fitted by applying a multi-parameter fitting routine to obtain the best fit based
on summation of Lorentzian peaks. We applied our mixing rule to first create the appropriate
alloy (Fig. \ref{fig:Schematic-of-mixing}(b)) with the particle diameters taken from \citep{MoskovitsJChemPhys02}.
The resulting theoretical behavior is shown as lines in Fig. \ref{fig:Moskovits}. The theory
again predicted quite well the shift in the position of the plasmon peak with varying alloy composition. 

In conclusion, we have developed a self consistent technique to determine/predict the effective
permittivity of ternary composites containing mixtures and/or alloy nanocomposites using the
Milton-Bergman lower bound expression for two component composites. We have successfully tested
our theory against previously studied experimental systems comprising mixtures of multi-metal
nanoparticles and/or alloys of Ag:Au:SiO$_{\text{2}}$, Ag-Cu:SiO$_{\text{2}}$, and Au-Cu:H$_{\text{2}}$O.
This approach predicts quite accurately the peak position and shift of plasmonic behavior in
these mixtures and alloys by utilizing experimentally available optical parameters \citep{SopraDataBase}
and without the need for any fitting. This result clearly indicates that our mixing approach
captures the essential physics of plasmonic behavior in these multi-component systems. This calculation
can readily be extended for nanocomposites with more than three components. However, at this
point, no experimental data exists to validate the calculations for such systems. In addition,
the shape of the plasmon peaks could also be fairly well reproduced by utilizing a simple correction
to the relaxation time of the electrons that accounts for the enhanced scattering at the metal
boundary as the particle size is reduced. This model allows for further corrections due to quantum
confinement effects as the particle size gets reduced below the 10 nm size scale \citep{halperin86}.
This approach could be of great value towards predicting optical properties in terms of plasmonic
behavior as a function of volume fraction, particle size and alloy composition in multi-component
mixtures and could guide the assembly of nanocomposites with tailored optical properties \citep{favazza06all,Bozhevolnyi06}. 

RK and RS acknowledge support by the National Science Foundation through grants \# DMI-0449258
and \# CTS-0335348 respectively.

\pagebreak

\section*{Figure captions}

\begin{enumerate}
\item Schematic of the binary tree approach to implement the self consistent binary mixing rule to
obtain the effective dielectric constant. (a) Application of the mixing rule to create a mixture
of two metals \emph{a} and \emph{b} in the host \emph{h}. (b) Application of the rule to create
an alloy nanocomposite in host \emph{h} from alloys of metals \emph{a} and \emph{b}. \label{fig:Schematic-of-mixing}
\item Comparison of theory and experiment for plasmonic behavior in a nanocomposite containing various
mixtures of Au and Cu in SiO$_{\text{2}}$. The theoretical results using the binary mixing rule
are presented as lines while experimental data (symbols) were extracted from the work of Magruder
et al. \citep{MagruderNonCrystSold94}. The peak position, shift and widths are predicted very
well using the mixing rule. The experimental data corresponds to various ratios of Ag:Cu achieved
by sequential ion implantation. \label{fig:Magruder}
\item Comparison of theory and experiment of plasmonic behavior in various Au-Cu alloy nanocomposites
in SiO$_{\text{2}}$ obtained by ion implantation and annealing. The experimental data (symbols)
was extracted from the work of Battaglin et al. \citep{BattaglinNucInsMethPhysB00} while the
theoretical calculations are represented as lines. Alloy compositions are indicated on the figure.
\label{fig:Battaglin}
\item Comparison of plasmonic behavior in colloidal solutions containing Au-Ag core-shell alloys for
two compositions. The symbols are experimental data extracted from Moskovits et al. \citep{MoskovitsJChemPhys02}
while lines are theory. The numbers within brackets in the legends represent the mole fraction
of Au in the colloidal solution.\label{fig:Moskovits}
\end{enumerate}
\pagebreak

\begin{figure}[tbh]
\subfigure[]{\includegraphics[height=2.25in,keepaspectratio]{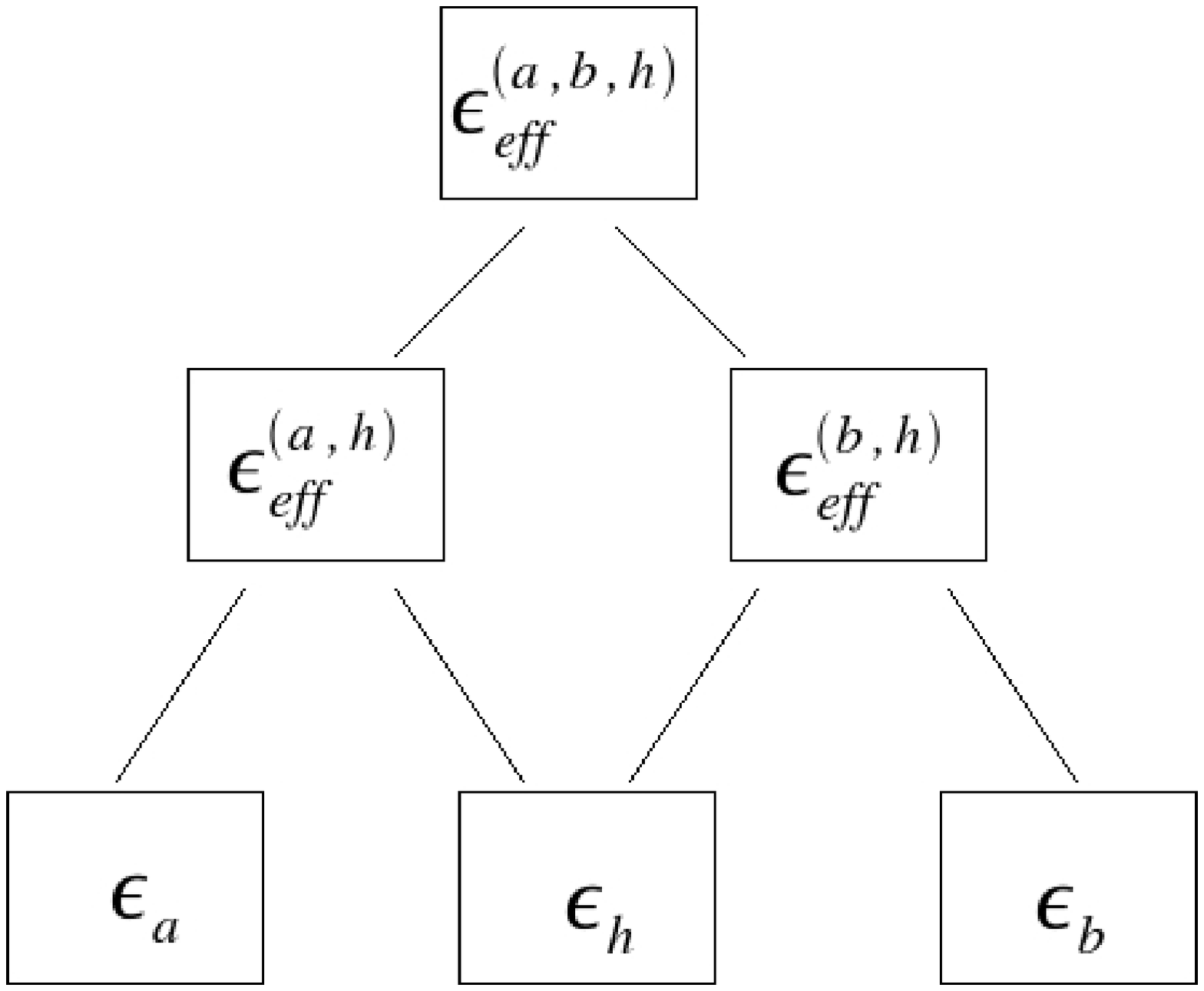}}~~\subfigure[]{\includegraphics[height=2.25in,keepaspectratio]{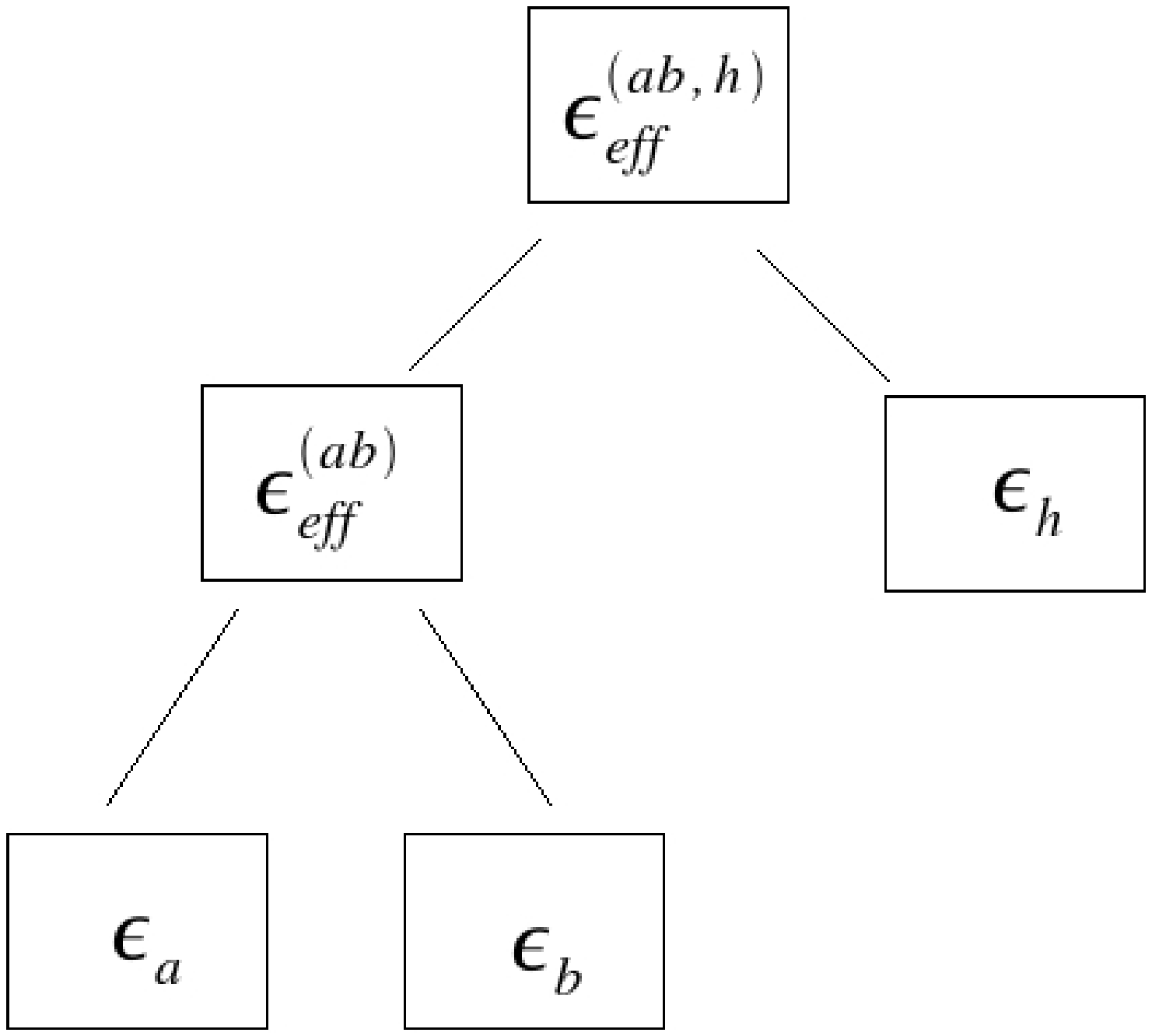}}

\caption{Schematic of the binary tree approach to implement the self consistent binary mixing rule to
obtain the effective dielectric constant. (a) Application of the mixing rule to create a mixture
of two metals \emph{a} and \emph{b} in the host \emph{h}. (b) Application of the rule to create
an alloy nanocomposite in host \emph{h} from alloys of metals \emph{a} and \emph{b}. \label{fig:Schematic-of-mixing}}
\end{figure}

\begin{figure}[tbh]
\begin{centering}\includegraphics[height=3in,keepaspectratio]{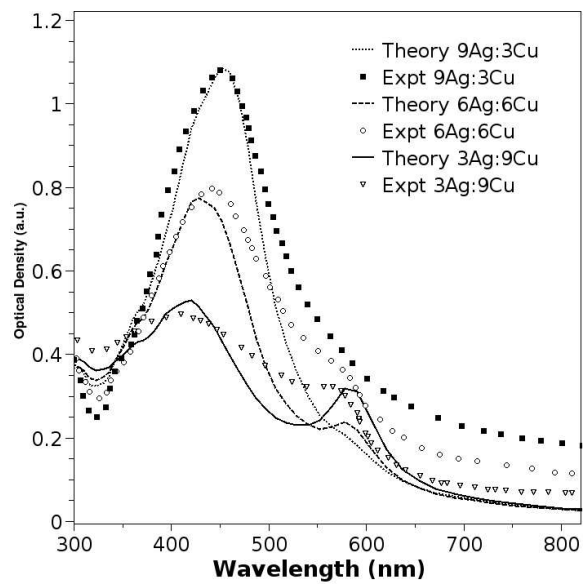}\par\end{centering}

\caption{Comparison of theory and experiment for plasmonic behavior in a nanocomposite containing various
mixtures of Au and Cu in SiO$_{\text{2}}$. The theoretical results using the binary mixing rule
are presented as lines while experimental data (symbols) were extracted from the work of Magruder
et al. \citep{MagruderNonCrystSold94}. The peak position, shift and widths are predicted very
well using the mixing rule. The experimental data corresponds to various ratios of Ag:Cu achieved
by sequential ion implantation. \label{fig:Magruder}}
\end{figure}

\pagebreak

\begin{figure}[tbh]
\begin{centering}\includegraphics[height=3in,keepaspectratio]{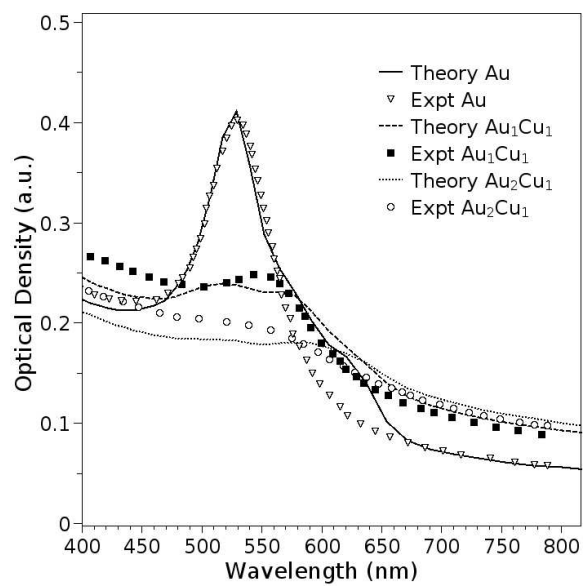}\par\end{centering}

\caption{Comparison of theory and experiment of plasmonic behavior in various Au-Cu alloy nanocomposites
in SiO$_{\text{2}}$ obtained by ion implantation and annealing. The experimental data (symbols)
was extracted from the work of Battaglin et al. \citep{BattaglinNucInsMethPhysB00} while the
theoretical calculations are represented as lines. Alloy compositions are indicated on the figure.
\label{fig:Battaglin}}
\end{figure}

\begin{figure}[tbh]
\begin{centering}\includegraphics[height=3in,keepaspectratio]{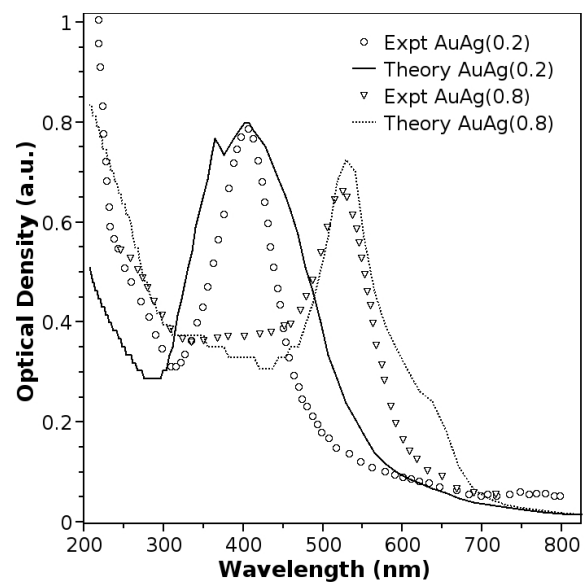}\par\end{centering}

\caption{Comparison of plasmonic behavior in colloidal solutions containing Au-Ag core-shell alloys for
two compositions. The symbols are experimental data extracted from Moskovits et al. \citep{MoskovitsJChemPhys02}
while lines are theory. The numbers within brackets in the legends represent the mole fraction
of Au in the colloidal solution.\label{fig:Moskovits}}
\end{figure}

\end{document}